\documentclass[conference]{IEEEtran}
\IEEEoverridecommandlockouts
\usepackage{amsmath}
\usepackage[linesnumbered,ruled,vlined]{algorithm2e}
\usepackage{algorithmic}
\usepackage{graphicx}
\usepackage[T1]{fontenc}

\usepackage[cmintegrals]{newtxmath}

\def\BibTeX{{\rm B\kern-.05em{\sc i\kern-.025em b}\kern-.08em
    T\kern-.1667em\lower.7ex\hbox{E}\kern-.125emX}}
\begin{document}

\title{Moment-Based Bound on Peak-to-Average Power Ratio and Reduction with Unitary Matrix
}

\author{\IEEEauthorblockN{Hirofumi Tsuda}
\IEEEauthorblockA{Department of Applied Mathematics and Physics\\
Graduate School of Informatics\\ Kyoto University\\
Kyoto, Japan \\
tsuda.hirofumi.38u@st.kyoto-u.ac.jp}
}

\maketitle

\begin{abstract}
Reducing Peak-to-Average Power Ratio (PAPR) is a significant task in OFDM systems. To evaluate the efficiency of PAPR-reducing methods, the complementary cumulative distribution function (CCDF) of PAPR is often used. In the situation where the central limit theorem can be applied, an approximate form of the CCDF has been obtained. On the other hand, in general situations, the bound of the CCDF has been obtained under some assumptions. In this paper, we derive the bound of the CCDF with no assumption about modulation schemes. Therefore, our bound can be applied with any codewords and that our bound is written with fourth moments of codewords. Further, we propose a method to reduce the bound with unitary matrices. With this method, it is shown that our bound is closely related to the CCDF of PAPR.
\end{abstract}


\section{Introduction}
Peak-to-Average Power Ratio (PAPR) is the ratio of the squared maximum amplitude to the average power. It is known that in-band distortion and out-of-band distortion are caused by a large input power since the output power is non-linear with respect to an input power for a large power regime \cite{concept}. It is known that these distortions reduce Signal-to-Noise Ratio (SNR) and the channel capacity \cite{ochiai}. Since signals with large PAPR tend to be distorted, it is demanded to reduce PAPR. Thus, many methods to reduce PAPR have been proposed \cite{complement}-\cite{timedomain}.

By definition, PAPR depends on a given codeword. However, PAPR is often regarded as a random variable since a codeword can also be regarded as a random variable. To investigate performances of PAPR-reducing methods, the complementary cumulative distribution function (CCDF) of PAPR is often evaluated \cite{slm}-\cite{pts}. Therefore, it is demanded to obtain the form of the CCDF. When each codeword is randomly and independently chosen from a given distribution and the central limit theorem can be applied, approximate forms of the CCDF have been obtained \cite{ochiai} \cite{extreme}. These results are based on that an OFDM signal can be regarded as a Gaussian process. Furthermore, it has been proven that usual coded-OFDM signals can be regarded as Gaussian processes \cite{convergence}. On the other hand, in the case where the central limit theorem cannot be applied, an approximate form has not been obtained.

In the case where the central limit theorem cannot be applied, the upper bounds of the CCDF of PAPR have been obtained \cite{exist}-\cite{litsyn}. It is expected to achieve lower PAPR as the upper bound decreases. Further, classes of error correction codes achieving low PAPR have been obtained \cite{exist}. To obtain upper bounds, some assumptions are often required. One of usual assumptions is about modulation schemes. Thus, with a given modulation scheme, methods to reduce PAPR have been discussed.

However, there is a case where codewords do not belong to a popular modulation scheme. For example, after applying an iterative clipping and filtering method \cite{iter} \cite{armstrong}, it is unclear what modulation scheme each symbol in codewords belongs to. In such a situation, known PAPR bounds could not be valid. Therefore, it is demanded to obtain a more generalized bound under no assumption about a modulation scheme.

In this paper, we derive an upper bound of the CCDF of PAPR with no assumption about a modulation scheme. Our bound is written in terms of fourth moments of codewords. As a similar bound, it has been proven that there is a bound which is written in terms of moments in BPSK systems \cite{general}. Therefore, our result can be regarded as a generalization of such an existing result.

To reduce PAPR, we apply the technique which has been developed in Independent Component Analysis (ICA) \cite{ica_ori}. The main idea of ICA is to find a suitable unitary matrix to reduce the kurtosis, which is a statistical quantity written in terms of fourth moments. From this idea used in ICA, it is expected that our bound can be reduced with unitary matrices since our bound is also written in terms of fourth moments of codewords. The known methods, a Partial Transmit Sequence (PTS) technique and a Selective Mapping (SLM) method are to modulate the phase of each symbol to reduce PAPR \cite{pts} \cite{slm}. Therefore, these known methods are to transform codewords with diagonal-unitary matrices and our method can be regarded as a generalization of these methods.

\section{OFDM System and PAPR}
In this section, we show the OFDM system model and the definition of PAPR. First, a complex baseband OFDM signal is written as \cite{ofdmcdma}
\begin{equation}
  s(t) = \sum_{k=1}^{K} A_k \exp\left(2 \pi j \frac{k-1}{T}t\right), \hspace{2mm} 0 \leq t < T,
  \label{eq:ofdm}
\end{equation}
where $A_k$ is a transmitted symbol, $K$ is the number of symbols, $j$ is the unit imaginary number, and $T$ is a duration of symbols. With Eq. (\ref{eq:ofdm}), a radio frequency (RF) OFDM signal is written as

\begin{equation}
  \begin{split}
    \zeta(t) &= \operatorname{Re}\{s(t)\exp(2 \pi j f_c t)\}\\
    &= \operatorname{Re}\left\{\sum_{k=1}^{K} A_k \exp\left(2 \pi j \left(\frac{k-1}{T} + f_c \right)t\right)\right\},
    \end{split}
\label{eq:baseband}
\end{equation}
where $\operatorname{Re}\{z\}$ is the real part of $z$, and $f_c$ is a carrier frequency. With RF signals, PAPR is defined as \cite{exist} \cite{compute}
\begin{equation}
\begin{split}
  & \operatorname{PAPR}(\mathbf{c})\\ 
= & \max_{0 \leq t < T}\frac{\left|\operatorname{Re}\left\{\displaystyle\sum_{k=1}^{K} A_k \exp\left(2 \pi j \left(\frac{k-1}{T} + f_c \right)t\right)\right\}\right|^2}{P_{\operatorname{av}}},
\end{split}
  \label{eq:PAPR}
\end{equation}
where $\mathbf{c} = (A_1,A_2,\ldots,A_K)^\top \in \mathcal{C}$ is a codeword, $\mathbf{x}^\top$ is the transpose of $\mathbf{x}$, $\mathcal{C}$ is the set of codewords, $P_{\operatorname{av}}$ corresponds to the average power of signals, $P_{\operatorname{av}} = \sum_{k=1}^{K}\operatorname{E}\{|A_k|^2\}$, and $\operatorname{E}\{X\}$ is the average of $X$. On the other hand, with baseband signals, Peak-to-Mean Envelope Power Ratio (PMEPR) is defined as \cite{exist} \cite{compute}
\begin{equation}
  \operatorname{PMEPR}(\mathbf{c}) = \max_{0 \leq t < T}\frac{\left|\displaystyle\sum_{k=1}^{K} A_k \exp\left(2 \pi j \frac{k-1}{T} t\right) \right|^2}{P_{\operatorname{av}}}.
  \label{eq:PMEPR}
\end{equation}
As seen in Eqs (\ref{eq:PAPR}) and (\ref{eq:PMEPR}), PAPR and PMEPR are determined by the codeword $\mathbf{c}$ and it is clear that $\operatorname{PAPR}(\mathbf{c}) \leq \operatorname{PMEPR}(\mathbf{c})$ for any codeword $\mathbf{c}$. In \cite{sharif}, it has been proven that the following relation is established under some conditions described below
\begin{equation}
  \left( 1 - \frac{\pi^2K^2}{2r^2}\right)\cdot \operatorname{PMEPR}(\mathbf{c}) \leq  \operatorname{PAPR}(\mathbf{c}) \leq  \operatorname{PMEPR}(\mathbf{c}),
  \label{eq:papr_pmepr}
\end{equation}
where $r$ is an integer such that $f_c = r/T$. The conditions that Eq. (\ref{eq:papr_pmepr}) holds are $K \ll r$ and $\exp(2 \pi j K/r) \approx 1$. In addition to these, another relation has been shown in \cite{litsyn}. From Eq. (\ref{eq:papr_pmepr}), PAPR is approximately equivalent to PMEPR for sufficiently large $f_c$. Throughout this paper, we assume that the carrier frequency $f_c$ is sufficiently large, and we consider PMEPR instead of PAPR. Note that this assumption is often used \cite{ochiai}.

\section{Bound of Peak-to-Average Power Ratio}
In this section, we show the bound of a CCDF of PAPR. As seen in Section II, PAPR and PMEPR depend on a given codeword. Since codewords are regarded as random variables, PAPR and PMEPR are also regarded as random variables. In what follows, we merely write $\operatorname{PAPR}$ in formulas when PAPR is a random variable. 

First, we make the following assumptions
\begin{itemize}
 \item the probability density of $\mathbf{c}$, $p(\mathbf{c})$ is given and fixed.
 \item the carrier frequency $f_c$ is sufficiently large.
 \item For $1 \leq k,l,m,n \leq K$, the statistical quantity $\operatorname{E}\{A_k A_l \overline{A}_m \overline{A}_n\}$ exists, where $\overline{z}$ is the conjugate of $z$.
\end{itemize}
The second assumption about a carrier frequency is often used \cite{ochiai}. As seen in Section II, PAPR is approximately equivalent to PMEPR if the carrier frequency $f_c$ is sufficiently large. Thus, we consider PMEPR instead of PAPR. The last assumption has been used in \cite{litsyn}. We call the quantity $\operatorname{E}\{A_k A_l \overline{A}_m \overline{A}_n\}$ the fourth moment of $A_k$, $A_l$ $\overline{A}_m$ and $\overline{A}_n$. For details about complex multivariate distributions and moments, we refer the reader to \cite{lapidoth} \cite{asymptotic_tech}. From the Cauchy-Schwarz inequality and this assumption, it can be proven that the average power $P_{\operatorname{av}}$ exists, that is, $P_{\operatorname{av}} < \infty$.

Let us consider the PAPR with a given codeword $\mathbf{c}=(A_1, A_2, \ldots, A_K)^\top$. In \cite{tellambura_bound}, the following relation has been proven
\begin{equation}
  \max_t|s(t)|^2 \leq \rho(0) + 2\sum_{i=1}^{K-1}|\rho(i)|,
  \label{eq:bound_env}
\end{equation}
where
\begin{equation}
  \rho(i) = \sum_{k=1}^{K-i}A_k\overline{A}_{k+i}.
\end{equation}
We let $\rho(K)$ be $0$. Note that the quantity $\rho(0)$ is the power of a codeword and that the time $t$ does not appear in the right hand side (r.h.s) of Eq. (\ref{eq:bound_env}). It is not straightforward to analyze Eq. (\ref{eq:bound_env}) since the absolute-value terms appear in Eq. (\ref{eq:bound_env}). To overcome this obstacle, we obtain the upper bound of r.h.s of Eq. (\ref{eq:bound_env}). From the Cauchy-Schwarz inequality, we obtain the following relation
\begin{equation}
\begin{split}
 \max_t|s(t)|^2  & \leq \rho(0) + 2\sum_{i=1}^{K-1}|\rho(i)|\\
 & \leq \sqrt{2K-1}\sqrt{|\rho(0)|^2 + 2\sum_{i=1}^{K-1}|\rho(i)|^2}.
\end{split}
\end{equation}
The above bound is rewritten as
\begin{equation}
 \max_t|s(t)|^4 \leq (2K-1)\left\{|\rho(0)|^2 + 2\sum_{i=1}^{K-1}|\rho(i)|^2\right\}.
\label{eq:bound_cauchy}
\end{equation}
The r.h.s of Eq. (\ref{eq:bound_cauchy}) is rewritten as
\begin{equation}
  \begin{split}
    & (2K-1)\left\{|\rho(0)|^2 + 2\sum_{k=1}^{K-1}|\rho(k)|^2\right\}\\
    =& (2K-1)\left\{\sum_{k=0}^{K-1}|\rho(k)|^2 + \sum_{k=0}^{K-1}|\rho(K-k)|^2 \right\}\\
    =&\frac{2K-1}{2}\left\{\sum_{k=0}^{K-1}|\rho(k) + \overline{\rho(K-k))}|^2\right.\\
    &\left. + \sum_{k=0}^{K-1}|\rho(k) - \overline{\rho(K-k)}|^2\right\}.
  \end{split}
    \label{eq:bound_expansion}
\end{equation}
From the above equations, the r.h.s of  Eq. (\ref{eq:bound_cauchy}) is written with periodic correlation terms and odd periodic correlation terms. These terms are written as
\begin{equation}
  \begin{split}
    \rho(k) + \overline{\rho(K-k)} &= \mathbf{c}^* B^{(k)}_{1,1}\mathbf{c},\\
    \rho(k) - \overline{\rho(K-k)} &= \mathbf{c}^* B^{(k)}_{-1,1}\mathbf{c},\\
  \end{split}
\end{equation}
where $\mathbf{z}^*$ is the conjugate transpose of $\mathbf{z}$, the matrices $B^{(k)}_{1,1}$ and $B^{(k)}_{-1,1}$ are
\begin{equation}
    B^{(k)}_{1,1} = \left( \begin{array}{cc}
                       O & I_{k}\\
                       I_{K-k} & O
                      \end{array} \right),\hspace{3mm}
   B^{(k)}_{-1,1} = \left( \begin{array}{cc}
                       O & -I_{k}\\
                       I_{K-k} & O
                             \end{array} \right).
\end{equation}
Since these matrices are regular, they can be transformed to diagonal matrices. From this general discussion, these matrices are decomposed with the eigenvalue decomposition as \cite{mypaper}
\begin{equation}
  \begin{split}
    B^{(k)}_{1,1} &= V^*D^{(k)}V \qquad B^{(k)}_{-1,1} = \hat{V}^*\hat{D}^{(k)}\hat{V},
  \end{split}
\end{equation}
where $V$ and $\hat{V}$ are unitary matrices whose $(m,n)$-th elements are
\begin{equation}
\begin{split}
V_{m,n} &= \frac{1}{\sqrt{K}}\exp\left(-2 \pi j \frac{mn}{K}\right),\\
 \hat{V}_{m,n} &= \frac{1}{\sqrt{K}}\exp\left(-2 \pi j n\left(\frac{m}{K} + \frac{1}{2K}\right)\right),
\end{split}
\end{equation}
and $D^{(k)}$ and $\hat{D}^{(k)}$ are diagonal matrices whose $n$-th diagonal elements are
\begin{equation}
\begin{split}
D_n^{(k)} &=  \exp\left(-2 \pi j k\frac{n}{K}\right),\\
  \hat{D}_n^{(k)} &=  \exp\left(-2 \pi j k\left(\frac{n}{K} + \frac{1}{2K}\right)\right).
\end{split}
\end{equation}
With these expressions, Eq. (\ref{eq:bound_cauchy}) is written as
\begin{equation}
  \max_t|s(t)|^4 \leq \frac{K(2K-1)}{2}\left\{\sum_{k=1}^{K}|\alpha_k|^4 + \sum_{k=1}^{K}|\beta_k|^4\right\},
\end{equation}
where $\alpha_k$ and $\beta_k$ are the $k$-th element of $\boldsymbol{\alpha}$ and $\boldsymbol{\beta}$ written as $\boldsymbol{\alpha}=V\mathbf{c}$ and $\boldsymbol{\beta} = \hat{V}\mathbf{c}$, respectively. With the codeword $\mathbf{c}$, the above inequality is written as
\begin{equation}
\begin{split}
 &  \max_t|s(t)|^4 \\
\leq & \frac{K(2K-1)}{2}\sum_{k=1}^K \left\{\left(\mathbf{c}^*V^*G_k V \mathbf{c}\right)^2 + \left(\mathbf{c}^*
      \hat{V}^* G_k \hat{V} \mathbf{c}\right)^2\right\},
\end{split}
  \label{eq:bound_expansion2}
\end{equation}
where $G_k$ is a matrix whose $(k,k)$-th element is unity and the other elements are zero. Note that $G_k^*G_k = G_k$. For the later convenience, we set $C_k = V^*G_k V$ and $\hat{C}_k = \hat{V}^*G_k \hat{V}$, respectively. Note that the matrices $C_k$ and $\hat{C}_k$ are positive semidefinite Hermitian matrices since $C_k$ and $\hat{C}_k$ are the Gram matrices. From Eq. (\ref{eq:bound_expansion2}), with a given codeword $\mathbf{c}$, the bound of the squared PAPR is obtained as

\begin{equation}
\begin{split}
 \operatorname{PAPR}(\mathbf{c})^2 &\leq \frac{\max_t|s(t)|^4}{P_{\operatorname{av}}^2}\\
&\leq \frac{K(2K-1)}{2P_{\operatorname{av}}^2}\sum_{k=1}^K \left\{\left(\mathbf{c}^* C_k \mathbf{c}\right)^2 + \left(\mathbf{c}^*
     \hat{C}_k \mathbf{c}\right)^2\right\}.
\end{split}
\end{equation}
In the above relations, the first inequality is obtained from the result that $\operatorname{PAPR}(\mathbf{c}) \leq \operatorname{PMEPR}(\mathbf{c})$.

From the above discussions, we have arrived at the bound of PAPR with a given codeword $\mathbf{c}$. From this bound, we can obtain the bound of the CCDF of PAPR as follows. Let $\operatorname{Pr}(\operatorname{PAPR} > \gamma)$ be the CCDF of PAPR, where $\gamma$ is positive. Then, the following relations are obtained
 \begin{equation}
  \begin{split}
  &\operatorname{Pr}(\operatorname{PAPR} > \gamma)\\
 = &\operatorname{Pr}(\operatorname{PAPR}^2 > \gamma^2)\\ 
\leq & \operatorname{Pr}(\max_t|s(t)|^4 > P_{\operatorname{av}}^2\gamma^2)\\
\leq & \frac{\operatorname{E}\left\{\max_t|s(t)|^4 \right\}}{P_{\operatorname{av}}^2\gamma^2}\\
\leq & \frac{K(2K-1)}{2P_{\operatorname{av}}^2 \gamma^2} \sum_{k=1}^K \operatorname{E}\left\{\left(\mathbf{c}^* C_k \mathbf{c}\right)^2 + \left(\mathbf{c}^* \hat{C}_k \mathbf{c}\right)^2\right\}.
  \end{split}
\label{eq:prob_bound}
 \end{equation}
In the course of deriving Eq. (\ref{eq:prob_bound}), the first equation has been obtained from the fact that PAPR is positive. The first inequality has been obtained from Eq. (\ref{eq:PMEPR}) and the fact that $\operatorname{PAPR}(\mathbf{c}) \leq \operatorname{PMEPR}(\mathbf{c})$ for any codeword $\mathbf{c}$ (see Section II). The second inequality has been obtained with the Markov inequality \cite{prob}. The last inequality has been obtained from Eq. (\ref{eq:bound_expansion2}). 

As seen in Eq. (\ref{eq:prob_bound}), the bound of the CCDF is written in terms of the fourth moments of codewords and the bound does not depend on a modulation scheme. Further, if each codeword $\mathbf{c}$ is randomly and uniformly chosen from the set of codewords $\mathcal{C}$ and the number of codewords is finite, then Eq. (\ref{eq:prob_bound}) is written as \cite{litsyn} 
 \begin{equation}
  \begin{split}
  &\operatorname{Pr}(\operatorname{PAPR} > \gamma)\\
\leq & \frac{1}{|\mathcal{C}|}  \frac{K(2K-1)}{2P_{\operatorname{av}}^2 \gamma^2} \sum_{\mathbf{c} \in \mathcal{C}} \sum_{k=1}^K \left\{\left(\mathbf{c}^* C_k \mathbf{c}\right)^2 + \left(\mathbf{c}^* \hat{C}_k \mathbf{c}\right)^2\right\},
  \end{split}
\label{eq:prob_unif_bound}
 \end{equation}
where $|\mathcal{C}|$ is the number of components in $\mathcal{C}$. 

\section{Reducing PAPR with Unitary Matrix}
In Section III, we have obtained the bound of the CCDF of PAPR. From Eq. (\ref{eq:prob_bound}), the bound is written in terms of the fourth moments of codewords. It is expected that PAPR decreases as the bound decreases. In this section, we propose a method to reduce the bound with unitary matrices. Our technique can be seen in ICA \cite{ica_ori} \cite{ica} since the main idea of ICA is to reduce the kurtosis, which is written in terms of the fourth moment.

In known methods, it has been proposed to modulate the phase of each symbol to reduce PAPR, and these methods are to transform a codeword with a diagonal-unitary matrix. Thus, our technique can be regarded as an extension of these methods.

In addition to the assumptions made in Section III, we make the following assumptions to introduce a technique to reduce our bound 
\begin{itemize}
 \item the number of components in codewords, $|C|$ is finite, that is, $|\mathcal{C}| = M < \infty$.
\item each codeword is chosen with equal probability from $\mathcal{C}$.
\end{itemize} 
Under the above assumptions, we propose a method to reduce PAPR with unitary matrices. The main idea of our method is to find unitary matrices which make our bound small. Through our technique, the average power $P_{\operatorname{av}}$ and SNR are preserved.

To introduce our method, we define subsets of codewords. First, from the above assumption, we can divide the codewords $\mathcal{C}$ into $N$ disjoint subsets which satisfy
\begin{equation}
 \mathcal{C} = \bigcup_{n=1}^N \mathcal{C}_n, \hspace{2mm} \mathcal{C}_m \cap \mathcal{C}_n = \emptyset \hspace{2mm}\mbox{for} \hspace{2mm}m \neq n. 
\end{equation}
Since the number of components in $\mathcal{C}$ is finite, each number of components in $\mathcal{C}_n$ is also finite. For each subset $\mathcal{C}_n$, we define a unitary matrix $W_n$. 

The scheme of our method is described as follows. First, let the transmitter and the receiver know the unitary matrices $\{W_n\}_{n=1}^N$. At the transmitter side, each codeword $\mathbf{c} \in \mathcal{C}_i$ is modulated to $W_i \mathbf{c}$ with the unitary matrix $W_i$. Then, the transmitter sends the number $i$ and $W_i \mathbf{c}$. At the receiver side, the symbol $\mathbf{y}$ and the number $i$ are received. Then, the receiver estimates the codeword $\hat{\mathbf{c}}$ as $\hat{\mathbf{c}} = W^*_i \mathbf{y}$. It is clear that $\hat{\mathbf{c}} = \mathbf{c}$ if $\mathbf{y} = W_i\mathbf{c}$. With the above scheme, the bound in Eq. (\ref{eq:prob_unif_bound}) is written as
 \begin{equation}
  \begin{split}
  &\operatorname{Pr}(\operatorname{PAPR} > \gamma)\\
\leq & \frac{1}{M}  \frac{K(2K-1)}{2P_{\operatorname{av}}^2 \gamma^2} \\
& \cdot \sum_{n=1}^N \sum_{\mathbf{c} \in \mathcal{C}_n} \sum_{k=1}^K \left\{\left(\mathbf{c}^* W_n^* C_k W_n \mathbf{c}\right)^2 + \left(\mathbf{c}^* W_n^* \hat{C}_k W_n \mathbf{c}\right)^2\right\}.
  \end{split}
\label{eq:unitary_bound}
 \end{equation}

In known methods, a PTS technique and a SLM method, one diagonal unitary matrix corresponds to one codeword. By contrast, in our methods, one unitary matrix corresponds to one set of codewords. This is the main difference between our method and the known methods.  

Let us consider the case where the channel is a Gaussian channel and the codeword $\mathbf{c} \in \mathcal{C}_i$ is sent. In such a situation, the received symbol $\mathbf{y}$ is written as 
\begin{equation}
 \mathbf{y} = W_i \mathbf{c} + \mathbf{n},
\end{equation}
where $\mathbf{n}$ is a noise vector whose components follow the complex Gaussian distribution independently. Then, the estimated codeword is written as
\begin{equation}
 \hat{\mathbf{c}} = \mathbf{c} + W_i^*\mathbf{n}.
\end{equation}
From the above equation, SNR is preserved through our method since the matrix $W_i$ is unitary. 

We have shown the main idea of our method. The remained problem is how to find $W_n$ which achieves low PAPR for $n=1,2,\ldots,N$. In our method, unitary matrices $W_n$ are given as the solutions which make our bound in Eq. (\ref{eq:unitary_bound}) small. To analyze our bound, we define 
\begin{equation}
\begin{split}
 &f\left(\left\{W_n\right\}_{n=1}^{N}\right)\\
 =& \sum_{n=1}^N \sum_{\mathbf{c} \in \mathcal{C}_n} \sum_{k=1}^K \left\{\left(\mathbf{c}^* W_n^* C_k W_n \mathbf{c}\right)^2 + \left(\mathbf{c}^* W_n^* \hat{C}_k W_n \mathbf{c}\right)^2 \right\}.
\end{split}
\label{eq:grad_function}
\end{equation}
Note that the variables of the function $f$ is the unitary matrices $\{W_n\}_{n=1}^N$ and that $f$ is a real function. To find the $\{W_n\}_{n=1}^N$ achieving low PAPR, we minimize $f\left(\left\{W_n\right\}_{n=1}^{N}\right)$ under the condition that $\{W_n\}_{n=1}^N$ is unitary. To minimize $f$, its gradient is necessary. However, in general, expressions involving complex conjugate or conjugate transpose do not satisfy the Cauchy-Riemann equations \cite{cookbook}. Thus, the function $f$ may not be differentiable. To avoid this, the generalized complex gradient of $f$ is defined as \cite{complex_gradient}
\begin{equation}
 \frac{\partial f}{\partial W_n} = \frac{\partial f}{\partial \operatorname{Re}\{W_n\}} + j\frac{\partial f}{\partial \operatorname{Im}\{W_n\}},
\end{equation}
where $\operatorname{Re}\{Z\}$ and $\operatorname{Im}\{Z\}$ are the real part and imaginary part of the matrix $Z$, respectively. With this definition, the gradient of $f$ with respect to $W_n$ is calculated as
\begin{equation}
\begin{split}
  &\frac{\partial f}{\partial W_n}\\
 =& 4 \sum_{\mathbf{c} \in \mathcal{C}_n} \sum_{k=1}^K \left\{\left(\mathbf{c}^* W^*_n C_k W_n \mathbf{c}\right)C_k  + \left(\mathbf{c}^* W^*_n \hat{C}_k W_n \mathbf{c}\right)\hat{C}_k \right\}W_n \mathbf{c} \mathbf{c}^*.
\end{split}
\label{eq:gradient}
\end{equation}
With the above equation, we propose the following gradient descent algorithm at the $l$-th iteration
\begin{equation}
 W^{(l+1)}_n \leftarrow W^{(l)}_n - \epsilon \frac{\partial f}{\partial W^{(l)}_n},
\end{equation}
where $W^{(l)}_n$ is the matrix obtained at the $l$-th iteration and $\epsilon$ is a positive parameter.
With the above iteration, we can obtain the matrix $W^{(l+1)}_n$ from $W^{(l)}_n$. Since the matrix $W^{(l+1)}_n$ is not always unitary, we have to project $W^{(l+1)}_n$ onto the set of unitary matrices. One method is to use the Gram-Schmidt process \cite{ica_app} \cite{ica} \cite{ica_complex}. First, we decompose the matrix $W^{(l+1)}_n$ as $W^{(l+1)}_n = (\mathbf{w}_1,\ldots,\mathbf{w}_K)^\top$ and update $\mathbf{w}_1 \leftarrow \mathbf{w}_1/\|\mathbf{w}_1\|_2$, where $\|\mathbf{z}\|_2$ is the $l_2$ norm of $\mathbf{z}$. Then, the following steps are iterated for $k=2,\ldots,K$
\begin{enumerate}
 \item $\mathbf{w}_k \leftarrow \mathbf{w}_k - \sum_{i=1}^{k-1}\mathbf{w}^*_k \mathbf{w}_i \mathbf{w}_i$.
\item $\mathbf{w}_k \leftarrow \mathbf{w}_k/\|\mathbf{w}_k\|_2$.
\end{enumerate}
Finally, the projected matrix is obtained as $W^{(l+1)}_n = (\mathbf{w}_1,\ldots,\mathbf{w}_K)^\top$.

With the above iterations, we can obtain a unitary matrix. However, it is unclear what order to choose and to normalize vectors. To avoid this ambiguity, a symmetric decorrelation technique has been proposed \cite{ica_app} \cite{ica} \cite{ica_complex}. A symmetric decorrelation technique is to normalize $W^{(l+1)}_i$ as
\begin{equation}
 W^{(l+1)}_n \leftarrow \left(W^{(l+1)}_n \left(W^{(l+1)}_n\right)^*\right)^{-1/2}W^{(l+1)}_n,
\label{eq:proj_unitary}
\end{equation}
where $(ZZ^*)^{-1/2}$ is obtained from the eigenvalue decomposition of $ZZ^* = F \Lambda F^*$ as $F\Lambda^{-1/2}F^*$ with $F$ being a unitary matrix, $\Lambda$ and $\Lambda^{-1/2}$ being diagonal positive matrices written as $\Lambda = \operatorname{diag}(\lambda_1,\lambda_2,\ldots,\lambda_K)$ and $\Lambda^{-1/2} = \operatorname{diag}(\lambda_1^{-1/2},\lambda_2^{-1/2},\ldots,\lambda_K^{-1/2})$, respectively. With the above projection, we can obtain the unitary matrix $W^{(l+1)}_n$. The algorithm of our method is summarized in Algorithm \ref{algo:unitary_method}.

\begin{algorithm}[htbp]
Set the initial unitary matrix $W^{(1)}_n$ for $n=1,2,\ldots,N$ and the iteration count $l=1$.\\
For $n=1,2,\ldots,N$, calculate $\frac{\partial f}{\partial W^{(l)}_n}$ and obtain the matrix $W^{(l+1)}_n$ as
\begin{equation*}
 W^{(l+1)}_n \leftarrow W^{(l)}_n - \epsilon \frac{\partial f}{\partial W^{(l)}_n}.
\end{equation*} \\
Project $W^{(l+1)}_n$ onto the set of unitary matrices for $n=1,2,\ldots,N$ with Gram-Schmidt process or Eq. (\ref{eq:proj_unitary}).\\
Let $\|W\|$ be the norm of the matrix $W$. If $\|W^{(l+1)}_n - W^{(l)}_n\| \approx 0$ for $n=1,2,\ldots,N$ or the iteration count $l$ gets sufficiently large, then stop. Otherwise, set $l \leftarrow l+1$ and go to step 2.
 \caption{How to Find Unitary Matrix in Our Method}
 \label{algo:unitary_method}
\end{algorithm}

\section{Numerical Results}
In this Section, we show the performance of our proposed method. As seen in Section II, we assume that the carrier frequency $f_c$ in Eqs (\ref{eq:baseband}) and (\ref{eq:PAPR}) is sufficiently large and then PAPR is approximately equivalent to PMEPR. Thus, we measure PMEPR instead of PAPR.  We set the parameters as $K=128$ and $M=2000$. To measure PMEPR, we choose oversampling parameter $J=16$ \cite{sharif} \cite{clip}. The modulation scheme is 16-QAM. All symbols are generated independently from the 16QAM set and then we obtain the set of codewords $\mathcal{C} = \left\{ \mathbf{c}_1,\ldots,\mathbf{c}_M \right\}$. In each $N$, we set the subsets of codewords as
\begin{equation}
 \mathcal{C}_n = \left\{ \mathbf{c}_{\frac{M}{N}(n-1)+1}, \mathbf{c}_{\frac{M}{N}(n-1)+2}, \ldots, \mathbf
{c}_{\frac{Mn}{N}} \right\}
\end{equation}
for $n=1,2,\ldots,N$. Thus, each subset of codewords is randomly obtained from the original 16QAM set and the number of components in $\mathcal{C}_n$ is $M/N$ for $n=1,2,\ldots,N$.
As initial points, we set $W_n^{(1)} = E$. where $E$ is identity matrix for $n=1,2,\ldots,N$. The gradient parameter $\epsilon$ is set as $\epsilon = N/M \cdot 1/K^2$. In Algorithm \ref{algo:unitary_method}, we have used the symmetric decorrelation technique described in Eq. (\ref{eq:proj_unitary}).

\begin{figure}[htbp]
\centering  
\includegraphics[width=3.0in]{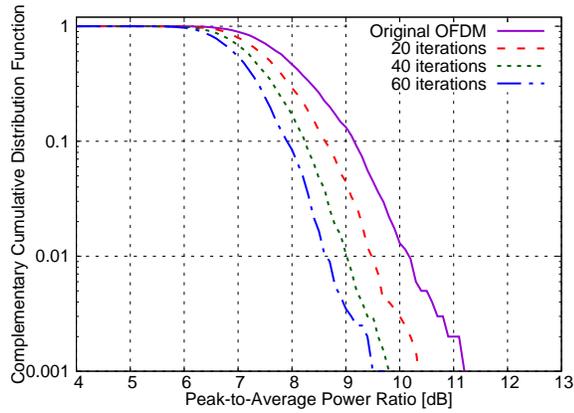}
\caption{PAPR in each iteration with $N=50$}
\label{fig:PAPR_iter_N50}
\end{figure}

Figure \ref{fig:PAPR_iter_N50} shows PAPR in our method with the parameter $N=50$. Each curve in the figure corresponds to the iteration times. As seen in this figure, PAPR gets small as the iteration time increases. This result shows that we can obtain the unitary matrices which achieve lower PAPR as the iteration time increases. Since our method is to reduce the bound of PAPR in Eq. (\ref{eq:unitary_bound}), this result implies that decreasing the our bound may lead to decrease PAPR. We conclude that our bound closely related to the CCDF of PAPR. 

\begin{figure}[htbp]
\centering  
\includegraphics[width=3.0in]{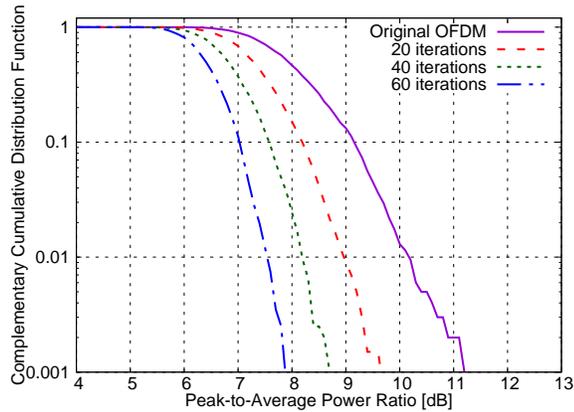}
\caption{PAPR in each iteration with $N=100$}
\label{fig:PAPR_iter_N100}
\end{figure}

Figure \ref{fig:PAPR_iter_N100} shows PAPR in our method with the parameter $N=100$. Similar to the result with the parameter $N=50$, our method achieves lower PAPR as the iteration time increases. However, from Fig. \ref{fig:PAPR_iter_N50} and Fig. \ref{fig:PAPR_iter_N100}, our method with the parameter $N=100$ achieves lower PAPR than one of our methods with $N=50$ at each iteration. The reason may be explained as follows. In our simulations, for $\mathbf{c} \in \mathcal{C}_n$, codewords are randomly and independently generated from 16QAM set and the average of $\mathbf{c}$ is $0$. Then, the average of the quantity $\mathbf{c}\mathbf{c}^*$ is
\begin{equation}
 \operatorname{E}\{\mathbf{c}\mathbf{c}^*\} = E,
\label{eq:assumption_id}
\end{equation}
where $E$ is the identity matrix. From the Cauchy-Schwarz inequality \cite{billingsley}, the lower bound of the bound in Eq. (\ref{eq:prob_bound}) can be written as

\begin{equation}
\begin{split}
& \frac{K(2K-1)}{2P_{\operatorname{av}}^2 \gamma^2} \sum_{k=1}^K \operatorname{E}\left\{\left(\mathbf{c}^* C_k \mathbf{c}\right)^2 + \left(\mathbf{c}^* \hat{C}_k \mathbf{c}\right)^2\right\}\\
\geq & \frac{K(2K-1)}{2P_{\operatorname{av}}^2 \gamma^2} \sum_{k=1}^K \operatorname{Tr}\left(\operatorname{E}\left\{C_k \mathbf{c} \mathbf{c}^*\right\}\right)^2 + \operatorname{Tr}\left(\operatorname{E}\left\{\hat{C}_k \mathbf{c} \mathbf{c}^*\right\}\right)^2\\
= & \frac{K(2K-1)}{2P_{\operatorname{av}}^2 \gamma^2} \sum_{k=1}^K \operatorname{Tr}\left(C_k\right)^2 + \operatorname{Tr}\left(\hat{C}_k\right)^2\\
= & \frac{K^2 (2K-1)}{P_{\operatorname{av}}^2 \gamma^2},
\end{split}
\label{eq:lower_bound}
\end{equation}
where $\operatorname{Tr}(X)$ is the trace of $X$. In the above inequalities, we have used $\operatorname{Tr}(C_k) = \operatorname{Tr}(\hat{C}_k) = 1$ and Eq. (\ref{eq:assumption_id}). It is clear that the above lower bound is invariant under the action $\mathbf{c} \mapsto W\mathbf{c}$, where $W$ is a unitary matrix. Let us consider the situations of the simulations with $N=50$ and $N=100$ and define the sample mean for each subset of codewords as
\begin{equation}
 g(\mathcal{C}_n) = \frac{1}{|\mathcal{C}_n|}\sum_{\mathbf{c} \in \mathcal{C}_n} \mathbf{c}\mathbf{c}^*.
\end{equation} 
Here, $|\mathcal{C}_n|$ is the number of components in the set $\mathcal{C}_n$, and we have assumed that $\mathbf{c}$ is randomly and independently chosen and that the quantity the average of $\mathbf{c}$ equals to $\mathbf{0}$ . Then, by the Law of Large Numbers, the quantity $g(\mathcal{C}_n)$ may be closer to the identity matrix as the number of components in $\mathcal{C}_n$ increases. From these discussions, if each subset $\mathcal{C}_n$ is randomly chosen from $\mathcal{C}$ and the number of components in $\mathcal{C}_n$ increases, then the quantity $g(\mathcal{C}_n)$ is nearly equivalent to the identity matrix. In such a situation, the lower bound in Eq. (\ref{eq:lower_bound}) may be tight. For these reasons, since each number of components in the subsets with $N=50$ is larger than one with $N=100$, our method with the parameter $N=100$ achieves lower PAPR than one of our methods with $N=50$ at each iteration.

\section{Conclusion}
In this paper, we have shown the bound of CCDF of PAPR and our proposed method to reduce PAPR. The main idea of our method is to transform each subset of codewords with the unitary matrix to reduce the bound of CCDF of PAPR. Further, the unitary matrices are obtained with the gradient method and the projecting method.

As seen in Section V, it may not be straightforward to reduce PAPR with our method when the quantity $g(\mathcal{C}_n)$ is nearly
 equivalent to the identity matrix. This obstacle may be overcome when we choose efficiently the subsets of codewords $\mathcal{C}_n$. Therefore, one of remained issues is to explore how to obtain the subsets of codewords $\mathcal{C}_n$. Further, it is necessary to explore other methods to reduce our bound.

\section*{Acknowledgment}
This work was partially supported by JSPS (KAKENHI) Grant Number 18J12903. The author would like to thank Dr. Shin-itiro Goto for his advise.

\end{document}